 \documentclass[manuscript,screen]{acmart}

\usepackage{tabularx}
\usepackage{booktabs}
\AtBeginDocument{%
  }

\setcopyright{acmlicensed}
\copyrightyear{2018}
\acmYear{2018}
\acmDOI{XXXXXXX.XXXXXXX}
\acmConference[Conference acronym 'XX]{Make sure to enter the correct
  conference title from your rights confirmation email}{June 03--05,
  2018}{Woodstock, NY}
\acmISBN{978-1-4503-XXXX-X/2018/06}

\begin{document}


\title{The UnScripted Trip: Fostering Policy Discussion on Future Human–Vehicle Collaboration in Autonomous Driving Through Design-Oriented Methods}





\author{Xinyan Yu}
\email{xinyan.yu@sydney.edu.au}
\orcid{0000-0001-8299-3381}
\affiliation{Design Lab,
  \institution{The University of Sydney}
  \city{Sydney}
  \state{NSW}
  \country{Australia}
}

\author{Julie Stephany Berrio Perez}
\email{stephany.berrioperez@sydney.edu.au}
\orcid{0000-0003-3126-7042}
\affiliation{The Australian Centre for Robotics,
  \institution{The University of Sydney}
  \city{Sydney}
  \state{NSW}
  \country{Australia}
}

\author{Marius Hoggenmueller}
\email{marius.hoggenmueller@sydney.edu.au}
\orcid{0000-0002-8893-5729}
\affiliation{Design Lab, Sydney School of Architecture, Design and Planning
  \institution{The University of Sydney} 
  \city{Sydney}
  \state{NSW}
  \country{Australia}
}

\author{Martin Tomitsch}
\email{Martin.Tomitsch@uts.edu.au}
\orcid{0000-0003-1998-2975}
\affiliation{Transdisciplinary School,
  \institution{University of Technology Sydney}
  \city{Sydney}
  \state{NSW}
  \country{Australia}
}

\author{Tram Thi Minh Tran}
\email{tram.tran@sydney.edu.au}
\orcid{0000-0002-4958-2465}
\affiliation{Design Lab, Sydney School of Architecture, Design and Planning
  \institution{The University of Sydney}
  \city{Sydney}
  \state{NSW}
  \country{Australia}
}

\author{Stewart Worrall}
\email{stewart.worrall@sydney.edu.au}
\orcid{0000-0001-7940-4742}
\affiliation{The Australian Centre for Robotics,
  \institution{The University of Sydney}
  \city{Sydney}
  \state{NSW}
  \country{Australia}
}

\author{Wendy Ju}
\email{wendyju@cornell.edu}
\orcid{0000-0002-3119-611X}
\affiliation{Jacobs Technion-Cornell Institute
  \institution{Cornell Tech} 
  \city{New York}
  \state{New York}
  \country{USA}
}

\renewcommand{\shortauthors}{Yu et al.}

\begin{abstract}
The rapid advancement of autonomous vehicle (AV) technologies is fundamentally reshaping paradigms of human–vehicle collaboration, raising not only an urgent need for innovative design solutions but also for policies that address corresponding broader tensions in society. To bridge the gap between HCI research and policy making, this workshop will bring together researchers and practitioners in the automotive community to explore AV policy directions through collaborative speculation on the future of AVs. We designed \emph{The UnScripted Trip}, a card game rooted in fictional narratives of autonomous mobility, to surface tensions around human–vehicle collaboration in future AV scenarios and to provoke critical reflections on design solutions and policy directions. Our goal is to provide an engaging, participatory space and method for automotive researchers, designers, and industry practitioners to collectively explore and shape the future of human–vehicle collaboration and its policy implications.
\end{abstract}




\maketitle


\section{Introduction \& Motivation}

With increasing levels of vehicle autonomy, models of human-vehicle collaboration are undergoing fundamental changes~\cite{XING2021103199}, challenging existing assumptions about control, responsibility, and interaction. These shifts not only require innovative design solutions to support effective collaboration, but also highlight the need for related policies to address the resulting ethical, social, and legal tensions~\cite{Bagloee2016Challenges, Daniel2015Preparing, Ten2024missing}. While the AutomotiveUI community has long contributed to the development of human–vehicle interfaces, user experience design, and AV safety, less attention has been given to how these insights translate into or inform the broader regulatory and policy landscape.

When addressing policy issues related to emerging technologies, where historical evidence is unavailable, policy considerations can lag behind and remain disconnected from human-computer interaction (HCI) work that centres on technology and its interaction with people~\cite{yang2024policyCollaboration}. For example, Uber's rapid introduction of its ride-sharing platform bypassed existing taxi regulations, forcing policymakers to respond only after the technology had already transformed urban transportation and user practices~\cite{Kessler2012Uber}. However, HCI researchers, with their cross-disciplinary expertise and critical perspectives, are particularly well-positioned to inform public policy~\cite{Spaa2019Boundaries, Lazar2010InteractingPolicy}. Increasingly, they are bringing their values and insights to policymaking processes, shaping debates around the societal and political impacts of digital technologies~\cite{Nathan2010InteractingwithPolicy, Roger2016PoliticsofMobility}. At the same time, practices in the design field such as futuring and speculative methods present abundant opportunities for HCI researchers to engage with policy. As noted by \citet{Spaa2019Boundaries}, these approaches are particularly valuable for envisioning alternative futures and supporting speculative policy development.

In responding to the calls for more collaborations across HCI-policy disciplinary boundaries~\cite{yang2024policyCollaboration,Davis2012USPolicymakers,Lazar2010InteractingPolicy}, this workshop proposes the use of a card game, a well-established approach in HCI for fostering constructive discussion around complex topics, to facilitate conversations on future AV policy directions. We designed \emph{The UnScripted Trip}, a card game rooted in fictional narratives of autonomous mobility, to surface tensions around human–vehicle collaboration in future AV scenarios and to provoke critical reflections on design solutions and policy directions.

\section{Workshop Goal \& Activity}
\subsection{Workshop Goal}

The goal of the workshop is to provide an engaging and participatory space for automotive researchers, designers, and industry practitioners to collectively speculate on the future of human-automation collaboration in autonomous driving and to explore potential policy directions. Through the construction of speculative narratives, participants are invited to surface and reflect on tensions that may emerge within new paradigms of human–vehicle collaboration. We hope this workshop will help bridge the gap in AV policy making by engaging perspectives from the AutomotiveUI community and increasing their participation in the policy process.

\subsection{\emph{The UnScripted Trip} Card Game}

Card games, also known as `games with a purpose'~\cite{von2008GameWithPurpose}, have become an effective tool for fostering constructive dialogue around challenging topics in HCI. For instance, \citet{Ballard2019JudgementCall} developed \emph{Judgement Call} to help product teams reflect on ethical considerations when designing AI-driven technologies, while \emph{What Could Go Wrong?} was developed by \citet{Martelaro2020WhatCouldGoWrong} to surface the potential downsides of autonomous vehicle technologies. 

Building on these approaches, we have developed and tested a card game called \emph{The UnScripted Trip}. Rooted in fictional narratives of autonomous mobility depicted in media, this card game is designed to surface tensions around human–vehicle collaboration in future scenarios and to provoke critical reflection on both design solutions and policy directions. The card game unfolds in two stages, as outlined below.

\subsubsection*{Stage 1. Construct a Backstory Narrative}
Participants construct a backstory narrative between a protagonist and a future AV by drawing one card from the \emph{Social Relationships} deck and one from the \emph{AV Functional Roles} deck. Each \emph{Social Relationship} card represents a speculative type of relationship an AV might form with humans---such as `sidekick' or `possessive partner'---derived from a review of fictional AV portrayals in fictional media previously conducted by the organisers. The \emph{AV Functional Roles} cards present a range of roles that the AV may assume, spanning from private-owned vehicles to Law
enforcement cars or retail pods. Using these two cards as a foundation, participants construct a backstory by imagining a fictional protagonist who operates or rides in the vehicle. They are prompted to imagine details such as the protagonist’s name, personality, and background, as well as the destination of their current trip by completing the prompt: \emph{`Today, they are on their way to…'}.

\subsubsection*{Stage 2. Trip of Tensions}
The fictional protagonist embarks on their trip to the designated destination, during which conflict situations will arise. These conflicts are collectively constructed through the use of a \emph{scenario} card deck and are further shaped by specific tensions between vehicle and human autonomy, determined by two spinning wheels. Each \emph{scenario} card represents a highly uncertain driving situation (e.g., encountering animals that suddenly appear on the highway), drawing on years of experience and research in the AV domain within the organising team, encompassing both engineering challenges and human–AV interaction perspectives. This approach aligns with the prevalence of scenario-based investigations within the AutomotiveUI community, as well as in recent workshops held at the conference~\cite{haoyu2023holistic, baby2023what, lee2022workshop}.

Then, participants spin one wheel to determine a Human Action (e.g., taking a nap) and another to determine an AV Response (e.g., request take-over), from which tensions or breakdowns may emerge.  Finally, participants reflect on the situation within the broader narrative they have created, speculating on design solutions that could resolve the tensions, and considering what policies might be needed to ensure the smooth implementation of these solutions.

\begin{figure}[H]
    \centering
    \includegraphics[width=0.9\linewidth]{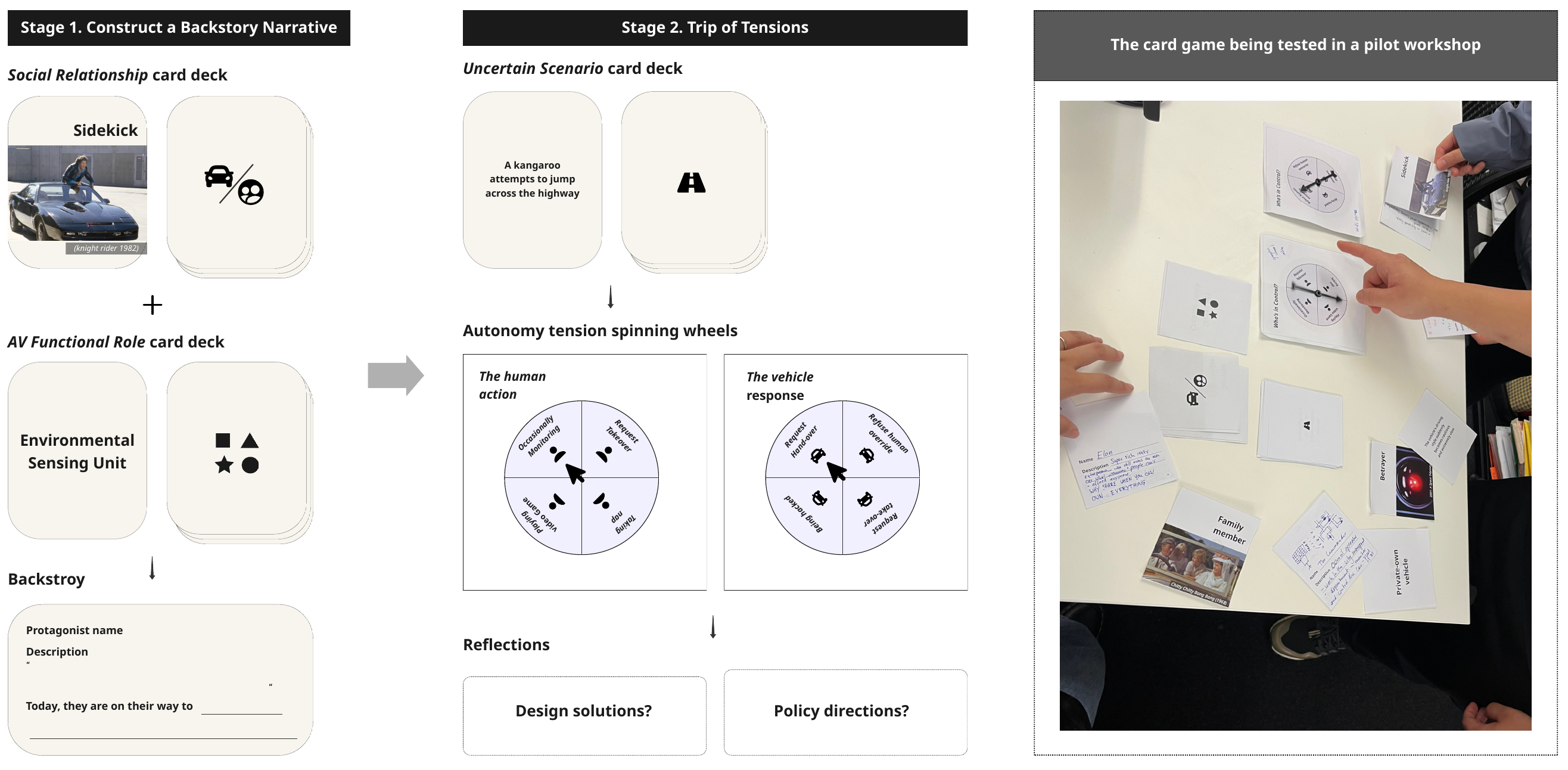}
    \caption{Overview of the card game \emph{The UnScripted Trip} and a photo taken during its pilot test.}
    \Description{}
    \label{fig:Internal}
\end{figure}




\section{Workshop Overview}

\subsection{Pre-Workshop Plan}

We will develop a dedicated website that provides detailed information about the workshop, serving as a centralised hub for disseminating updates and related materials. To ensure we reach a broad audience, we will share the workshop’s call for participants across social media platforms such as BlueSky and LinkedIn, and leverage our professional networks. 

The workshop will not require the submission of position papers. Instead, we will encourage participants to prepare and bring an example of an autonomous vehicle portrayed in film or other media to share with the group, which helps set the stage for our fictional media-inspired card game and form part of the ice-breaker activity.

\subsection{Workshop Schedule}

We propose a half-day workshop (3.5 hours, including a 30-minute break) that will accommodate approximately 20-30 attendees, including the organisers. Below, we outline a tentative schedule for the workshop day:

\begin{itemize}
    \item \textbf{Opening \& Introduction (15 mins)}: The workshop will begin with the organisers welcoming participants and providing a brief overview of the workshop’s aims, agenda, and expected outcomes. This will be followed by ice-breaker activities, where participants introduce themselves and share an example of an AV from film or other media, as requested as the pre-workshop preparation.
    
     \item \textbf{Keynote Talk (30 mins)}: To set the tone for the workshop, we will invite a keynote speaker from the automotive industry to provide insights into current trends, challenges, and opportunities in AV development and regulation. These insights will help ground subsequent discussions in real-world contexts and emerging issues.
      \item \textbf{Card Game Activity (70 mins)}: After introducing the card game mechanics, participants will break into pairs to complete the first phase of backstory narrative construction. Next, these pairs will join together to form larger groups of 4–6 and select one backstory to carry forward into the second phase. Members of the organising team will join and facilitate each group during this phase. After the game, participants present their design solutions and policy directions for addressing the conflicts to the larger group.
 \item \textbf{After-activity Discussion (45 mins)}: After the card game activity, participants will come together for a group discussion to reflect on policy directions for addressing uncertainty in AV driving, as well as their experiences with and suggestions on the card game method itself.

\end{itemize}

\begin{table}[h]
  \small
  \caption{Tentative schedule activities of the workshop}
  \label{tab:schedule}
  \begin{tabular}{l | l | l}
    \toprule
    \textbf{Activity} &\textbf{Duration} & \textbf{Time} \\
    \midrule
    Opening \& Introduction & 15 min & 13:45 - 14:00\\ 
    \textbf{Keynote Talk} & 30 min & 14:00 - 14:30 \\ 
    \textbf{The UnScripted Trip Card Game} & 75 min & 14:30 - 15:45\\ 
    Coffee Break & 30 min & 15:45 - 16:15 \\
    \textbf{Post-activity Discussion} & 45 min & 16:15 - 17:00 \\ 
    Wrapping-up \& Next Steps & 30 min &  17:00 - 17:30 \\
  \bottomrule
\end{tabular}
\end{table}

\subsection{Post-Workshop Plan and Expected Outcomes}
\begin{itemize}
    \item \textbf{Workshop Report:} The workshop activities and participants’ discussions on AV policy directions will be synthesised into a workshop report. We aim to publish the report through relevant venues in the automotive and HCI field.
    
    \item \textbf{Card Game Evaluation and Refinement:} We will review the card game activities and collect formal feedback from participants through a post-workshop survey. Our aim is to understand how well the game mechanics facilitate discussion, the breadth and depth of the issues explored, and the overall engagement of the activity. Insights and feedback gathered during the workshop will be used to refine the \emph{The UnScripted Trip} card game for future use and adaptation to broader contexts.
    
\end{itemize}

\section{Organisers}
\textbf{Xinyan Yu} is a PhD candidate in the Design Lab at the University of Sydney's School of Architecture, Design, and Planning. Her research centres around human-robot collaboration in urban settings, exploring bystanders' pro-social behaviours towards robots.

\textbf{Dr. Julie Stephany Berrio} holds the position of Research Associate at the Australian Centre for Robotics, which is affiliated with the University of Sydney. Her research is primarily centred around perception, mapping, and digital twins, specifically focusing on their application in vehicular technology.

\textbf{Dr. Marius Hoggenmüller} is a Lecturer in Interaction Design in the Design Lab at the University of Sydney's School of Architecture, Design, and Planning. His work focuses on prototyping interactions with emerging technologies in cities, such as urban robots and autonomous systems.

\textbf{Dr. Martin Tomitsch} is a Professor and Head of the Transdisciplinary School at the University of Technology Sydney, and a founding member of the Media Architecture Institute, the Urban Interfaces Lab, and the Life-centred Design Collective. He works at the intersection of design and technology with a focus on cities and responsible innovation.

\textbf{Dr. Tram Thi Minh Tran} is a postdoctoral researcher at the University of Sydney, Australia. She holds both her Master’s and PhD degrees from the same university. Her research explores the applications and implications of emerging technologies, with a particular focus on AR/VR and autonomous mobility.

\textbf{Dr. Stewart Worrall} is a Senior Research Fellow, and leads the Intelligent Transportation Systems group at the Australian Centre for Robotics which is part of the University of Sydney. His research aims to track and predict the intentions of drivers and pedestrians, and better understand how this can be used to improve the way that vehicles interact with people.

\textbf{Dr. Wendy Ju} is an Associate Professor at the Jacobs Technion-Cornell Institute at Cornell Tech and the Technion. Her work in the areas of human-robot interaction and automated vehicle interfaces highlights the ways that interactive devices can communicate and engage people without interrupting or intruding.


\bibliographystyle{ACM-Reference-Format}
\bibliography{sample-base}


\end{document}